\documentclass[journal,twocolumn,10pt]{IEEEtran}
\usepackage{amssymb}
\usepackage{cite}
\usepackage{graphicx}
\usepackage{psfrag}
\usepackage{subfigure}
\usepackage{url}
\usepackage{stfloats}
\usepackage{amsmath}
\usepackage{float}
\usepackage{bm}

\usepackage{mathrsfs}
\usepackage{subeqnarray}
\usepackage{cases}
\usepackage{stfloats}
\usepackage{subeqnarray}
\usepackage{cases}

\usepackage{algorithm}
\usepackage{algorithmic}

\usepackage{xcolor}
\usepackage{soul}

\usepackage{amsthm}

\theoremstyle{plain}

\theoremstyle{definition}

\theoremstyle{remark}

\definecolor{txcolor}{rgb}{0.8,0.09,0.3}

\ifCLASSINFOpdf
\else
\fi

\allowdisplaybreaks[3]

\begin{document}
\title{Mobile Service-Based Cooperative Scheduling for High-Mobility Vehicular Networks}

\author{Yu~Zhang, ~Ke~Xiong,~\IEEEmembership{Member, ~IEEE,} Pingyi Fan,~\IEEEmembership{Senior Member, ~IEEE}, Xianwei Zhou\\

\thanks{Manuscript received xxx, 2015; revised xxx,  and
accepted xxx. Date of publication June xxx;
date of current version xxx. This work was supported by the National
Basic Research Program of China (973 Program), no. 2012CB316100(2) and also by the Conhealth Project, no. 294923.

Yu Zhang and Xianwei Zhou are with School of Computer and Communication Engineering, University of Science and Technology Beijing, P.R. China, 100083. e-mail:  zhangyu.ustb.scce@foxmail.com, xwzhouli@sina.com.

Ke~Xiong is with the School of Computer and Information Technology,  Beijing Jiaotong University,  Beijing,  R.P. China, 100084 and  the Department
of Electronics Engineering,  Tsinghua University, Beijing,  R.P. China  100044. e-mail:  kxiong@bjtu.edu.cn.

Pingyi Fan is with the Department of Electronics Engineering,  Tsinghua University,  Beijing,  R.P. China, 100084. e-mail:  fpy@tsinghua.edu.cn.
}
}
\markboth{IEEE Transactions on Intelligent Transportation Systems, ~Vol.~x, No.~x, x.~2015}%
{Shell \MakeLowercase{\textit{et al.}}: Bare Demo of IEEEtran.cls for Journals}


\maketitle

\begin{abstract}
This paper investigates the downlink scheduling for relay-aided high-mobility vehicular networks, where the vehicles with good vehicle-to-infrastructure (V2I) links are employed as cooperative relay nodes to help the ones with poor V2I links forward information via vehicle-to-vehicle (V2V) links. In existing works, instantaneous achievable information rate was widely adopted to perform the link scheduling, but it is not efficient for vehicular networks, especially for high-mobility scenarios. Different from them, in this paper, we introduce the mobile service to describe the mobile link capacity of vehicular networks and then we propose a mobile service based relaying scheduling (MSRS) for high mobility vehicular networks.
In order to explore the system information transmission performance limit, we formulate an optimization problem to maximize the mobile service amount of MSRS by jointly scheduling the V2I and V2V links. Since it is a combinational optimization problem which is too complex to solve, we design an efficient algorithm with low-complexity for it, where Sort-then-Select, Hungarian algorithm and Bisection search are employed. Simulation results demonstrate that our proposed MSRS is able to achieve the optimal results with an optimal approximation ratio larger than 96.5\%. It is also shown that our proposed MSRS is much more efficient for high-mobility vehicular systems, which can improve the system average throughput with increment of 3.63\% compared with existing instantaneous achievable information rate based scheduling method, and with  15\% increment compared with traditional non-cooperation scheduling method, respectively.
\end{abstract}


%\begin{IEEEkeywords}
%IEEEtran, journal, \LaTeX, paper, template.
%\end{IEEEkeywords}

% \fi
\IEEEpeerreviewmaketitle

\section{Introduction}
\subsection{Backgaround}
Nowadays, Intelligent transportation systems (ITS) become more and more important in people's daily life, because the fast development of urbanization enables people to spend more and more time within moving vehicles\cite{R:IIS1}\cite{R:IIS2}. To support various information services for ITS including on board video in vehicles and real-time transport
information dissemination, ever-increasing data are required to be delivered over vehicular networks\cite{R:VN1}\cite{R:VN2}.

In vehicle networks, there are generally two types of communication paradigms, i.e., vehicle-to-infrastructure (V2I) and vehicle-to-vehicle (V2V) communications\cite{R:HV1}\cite{R:HV2}, where V2I communications provide the connection between vehicles and the Internet via roadside base stations (BS), which can support  real-time applications, e.g., online music radio, and V2V communications provide the connection between nearby vehicles via dedicated short-range communications (DSRC) technologies, which can support active safety applications, e.g., variable message signs and ramp signalling on motor ways. Since vehicular networks often suffer from performance deterioration caused by the instability of mobile wireless channel and the variation of network topology, the integration of V2I communication and V2V communication in a single vehicular system  recently has attracted more attention \cite{R:HV1}-\cite{R:CV2}.

\subsection{Related work}
In order to further improve system performance of vehicular networks, a variety of advanced communication technologies were employed. Among them, cooperative relaying is widely considered as one of the most promising technologies, as it is capable of enhancing the system performance by exploiting transmission diversity offered by relay nodes \cite{R:CV2}\cite{R:CC1}.

With cooperative relaying, the signal attenuation caused by the long distance between the vehicle and BS can be released and more vehicular users can be embraced in the service area with more favorable channel conditions, which consequently yields better communication quality and higher system throughput. To well support cooperative relaying transmissions in vehicle networks, several relaying protocols including amplify-and-forward (AF) and decode-and-forward (DF) were employed, see e.g., \cite{R:AF2}-\cite{R:DF1}. In \cite{R:AF2}, it investigated the performance of a V2V cooperative AF relaying scheme for inter-vehicle communications, where optimal power allocation was studied to enhance system performance. In \cite{R:AF1}, the AF cooperative relaying was employed in downlink vehicular networks and pairwise error probability expression and the achievable diversity gains were analyzed. In \cite{R:DF1},  it analyzed the performance of a relay selection scheme for cooperative vehicular networks with the DF relaying, where only the ``best'' relay was selected and the outage performance was discussed.

In relay-aided cooperative vehicular systems, different vehicles have different communication performances over V2I and V2V links so that link scheduling becomes very essential\cite{R:F1}\cite{R:F2}. For example, in a  multiple vehicular user system, how to determine which vehicles should be scheduled to communicate with the BS over the V2I links and which ones should be scheduled to communicate with BS over the two-hop V2I-V2V links may greatly affect the system performance. Moreover, how to establish effective cooperative strategy among the vehicles also has significant impacts on system performance. Thus, some works focused on the link scheduling for relay-assisted vehicular networks, see e.g. \cite{R:S1}\cite{R:S2}. In \cite{R:S1}, it proposed graph-based scheduling to achieve the maximum sum rate for vehicular networks, where DF protocol was considered. In \cite{R:S2}, it proposed a scheduling scheme based on the multi-choice knapsack problem to jointly optimize the link scheduling and radio resource allocation for V2V links.

\subsection{Motivations}
However, all existing works scheduled the V2I and V2V links of vehicular networks on the basis of achievable instantaneous information rate of the links. As is known, to capture the dynamic change of network topology and link quality, link scheduling should be performed periodically. That is, for a given period $T$, all links are scheduled in terms of the link state information at the beginning of $T$, i.e., at time $t$, which lasts the whole time interval of $T$. In relatively low mobility scenarios, the link state and network topology can be regarded to be constant within each $T$, so that current achievable instantaneous information rate based scheduling is efficient for $T$ although it is calculated in terms of the instantaneous rates of all links at $t$. But in relatively high mobility scenarios, the link state and network topology may change greatly even for a small time interval $\Delta t$ ($\Delta t \leq T$). Therefore, the achievable instantaneous information rate of the link obtained at time $t$ may be very different from that at $t+\Delta t$. In this case, if we still use the achievable instantaneous information rate at instant $t$ to perform the link scheduling over its later period of $T$, it may be improper and even result in a great system performance loss.

Motivated by this, we focus on the link scheduling scheme design for relatively high mobility vehicular networks in this paper. Such a high mobility consideration is significant  because high-way transportation has been widely deployed all over the world, where the total length of the high ways in China has exceeded $10 \times 10^3$km with the maximum speed up to 120km/h  and that in USA has exceeded 88730km with the maximum speed up to 137km/h in some states.

\subsection{Contributions}\label{Sec:CC}
We consider a heterogenous vehicular system, where the V2I link and the V2V link  adopt different wireless protocols and utilize different communication resources. Specifically, V2I link delivers information via the Third-Generation Long-Term Evolution advanced version (LTE-A) and V2V link delivers information via the DSRC protocol. Since compared with the uplink communication, the downlink communication in which the BS transmits Internet data to the vehicular nodes (VNs), has much larger volumes of data to deliver, we investigate the downlink scheduling for the system.

The contributions of this paper are summarized as follows.

\emph{Firstly}, we introduce the mobile service amount to characterize the information transmission capacity of high-mobility V2V and V2I links. Based on this, we investigate the optimal relaying scheduling scheme to maximize the system total mobile service amount. Our proposed mobile service based relaying scheduling (MSRS) is very different from existing achievable instantaneous information rate based relaying scheduling (IRRS) which aimed to maximize the sum rate of vehicular systems. Since mobile service is the accumulating information transmission amount over a given time period $T$, it can adapt to the fast variation of the link state and the fast change of vehicular network topology. Thus, our proposed MSRS is more efficient for high mobility vehicular networks.
\begin{figure}
\centering
\includegraphics[width=0.49\textwidth]{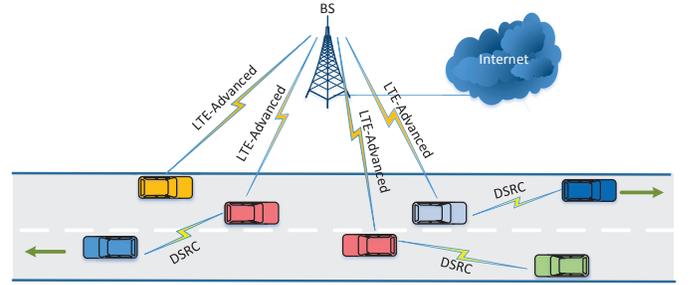}
\caption{Heterogenous vehicular network model composed of V2I and V2V links}\label{Fig:model}
\end{figure}

\emph{Secondly}, to explore the system performance limit of our proposed MSRS, we formulate an optimization problem for it. Since the problem is too complex to solve, we design an efficient algorithm with low-complexity. Specifically, the VNs with high V2I mobile service amount are scheduled as helping relaying vehicles (RV) and those with low V2V mobile service amount are scheduled as aided vehicles (AV). For a given number of AV, we construct a square benefit matrix and then adopt Hungarian Algorithm to calculate the optimal pairing between the RVs and the AVs. Then we use the Binary search to update the number of AVs to get the optimal solution.

\emph{Thirdly}, we provide extensive simulation results to compare and discuss the system performance. It shows that, our proposed low-complexity MSRS algorithm can achieve the approximating maximum mobile service amount with a probability larger than 96.5\% for vehicular networks. It is also shown that our proposed MSRS is much more efficient for high-mobility scenarios, which enhances the system performance with 3.63\% increment, compared with existing IRRS and 15\% increment compared with traditional non-cooperation scheduling method, respectively.

The rest of the paper is organized as follows. In Section \ref{Sec:SectII}, it describes the system model.
Section \ref{Sec:SectIII} formulates an optimization problem for our proposed MSRS. In Section \ref{Sec:SectIV}, we design low-complex algorithm for MSRS and provides extensive simulation results in Section \ref{Sec:SectV}. Finally we summarizes the paper with some conclusions in Section \ref{Sec:SectVI}.

\section{System Model} \label{Sec:SectII}
\subsection{Network Model}

Consider a vehicular communication system as shown in Figure \ref{Fig:model}, where $N$ moving VNs on the road desires to receive data from a common BS located on the roadside.  Similar to normal cellular users in LTE-A networks, VNs can communicate with the BS via direct V2I links between the BS and each VN. Due to large propagation path loss, only the VNs relatively close to the BS are able to acquire good communication service and the ones relatively far away from the BS often experience poor communication service. Heterogenous V2I and V2V links are considered, where VNs communicate with each other via DSRC V2V connections while V2I links employ the LTE-A.

To improve the communication capacity for the distant VNs (i.e., the aforementioned AV in Section \ref{Sec:CC}), the VNs with high communication capacity to the BS are scheduled as relays (i.e., the aforementioned RV in Section \ref{Sec:CC}) to help the BS to forward data to them. That is, the RV first receives the data from the BS via a LTE-A V2I link and then forwards the received data to the AV via a DSRC V2V link. With such a relaying strategy,  the information transmission capacity between the BS and the distant VN can be enhanced. As a result, the total throughput of the whole vehicular network is improved.

In such a system, for RV, it receives data from the BS through the LTE-A link and forwards data through the DSRC link to its AV. Note that RV is also a user of the BS and it also receives information from the BS for itself. For AV, it receives data over the two-hop V2I-V2V relaying path. Actually, in Figure \ref{Fig:model} it can be seen that, besides RV and AV, there is another kind of VNs, which just receives information from the BS for themselves without helping any AV. We refer them as common vehicle (CV).

\subsection{Channel Model}
It is known that signal strength is the key factor to wireless communication systems, where a strong received signal may yield a high received signal-to-noise-ratio (SNR), consequently a high transmission rate, and a weak received signal may lead to a low received SNR and a low transmission rate.

In vehicular networks, especially in suburban scenarios and viaduct scenarios, scatterers are very sparse,
the effect of large-scale fading (propagation path loss) generally is much more serious than that of small-scale fading. Similar to many existing work, see e.g., \cite{R:CHUANG},  we ignore the channel variation caused by small-scale fading and assume that the change of received signal strength only depends on
the position shifts of vehicles. In this case, the path
loss can be expressed as a function of the distance between the transmitter and the receiver as
\begin{equation}
L(d(t)) = L(d_0) + 10\alpha {\rm log}_{10}\bigg(\frac{d(t)}{d_0}\bigg) \rm{dB},
\end{equation}
where $L(d_0)$ is the path-loss
attenuation at reference distance $d_0$,
which is affected by the carrier frequency, the heights of
the transmitter and receiver antennas, different climate or
geology conditions, etc. $\alpha$ is the
path-loss exponent (usually $\alpha\geq 2$) and $d(t)$ is the distance between the transmitter and the
receiver at time $t$. As $L(d(t)) $ also can be expressed by
$
L(d(t)) = 10{\rm log}_{10}(\frac{P_s}{P_r}).
$
For a given transmit power $P_s$, the received power is
\begin{equation}
P_r=\frac{P_s}{10^{\frac{L(d(t))}{10}}}.
\end{equation}

\subsection{Instantaneous Achievable Information Rate of the Links}
Assuming that all links in the system are additive white Gaussian noise (AWGN) channel with a reference noise power of $\sigma_0^2$. Therefore, at
the moment $t$, the instantaneous achievable information rate between a transmit and its receiver is
\begin{equation}\label{Eq:capacity}
C(t) = {\rm log}_{2}\bigg(1+\frac{P_r}{\sigma_0^2}\bigg)= {\rm log}_{2}\bigg(1+\frac{P_s}{\sigma_0^2 10^{\frac{L(d(t))}{10}}}\bigg).
\end{equation}

\subsubsection{Direct Communications
} 
\begin{figure}
\centering
\includegraphics[width=0.475\textwidth]{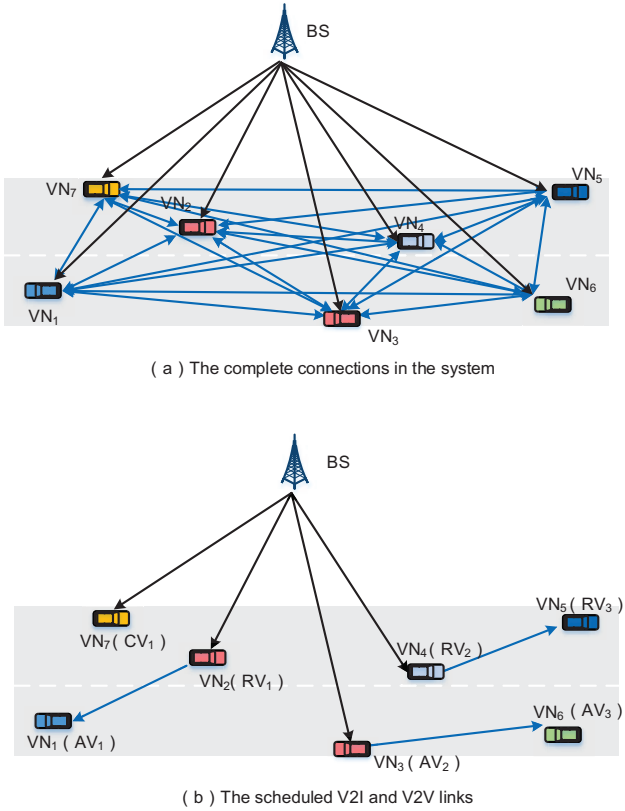}
\caption{Illustration of link scheduling in vehicular networks}\label{Fig:Links}
\end{figure}

For V2I link, it connects the BS and RV or the BS and CV. In infrastructure-based LTE-A system, the communications over the V2I link is based on the LTE-Advanced specification\cite{R:3GPP}, where the entire system LTE radio
resources are divided into resource blocks (RBs) along the
time and/or frequency domain. Let $K_{\rm{LTE}}$ be the number of the total available LTE RBs in the system. All VNs share the $K_{\rm{LTE}}$ RBs to communicate with the BS as shown by the black lines in Figure \ref{Fig:Links}(a). For each VN, it is assigned with $\lfloor \frac{K_{\rm LTE}}{N} \rfloor$ RBs. Thus, the achievable information rate that the $i$th VN can enjoy from the BS is given by
\begin{equation}\label{Eq:V2I}
C_{i,\rm{B}}(t) = \bigg\lfloor \frac{K_{\rm{LTE}}}{N} \bigg\rfloor {\rm log}_{2}\bigg(1+\frac{P_{\rm{B}}}{\sigma_I^2 10^{\frac{L_{i,\rm{B}}(d_{i,\rm{B}}(t))}{10}}}\bigg),
\end{equation}
where $L_{i,\rm{B}}$, $d_{i,\rm{B}}(t)$, $P_{\rm{B}}$ and $\sigma_I^2$ are the path-loss, the distance between the $i$th VN and the BS at time $t$,  the transmit power of the BS and the AWGN noise power over V2I link of each RB, respectively.

For V2V link, it connects the VNs through DSRC links, following the IEEE 802.11p specification \cite{R:DSRC}. Let $K_{\rm{DSRC}}$ and $n_{\textmd{\tiny AV}}$ be the number of the total available DSRC RBs and the number of AVs in the system. Thus, all $n_{\textmd{\tiny AV}}$ AVs share the $K_{\rm{DSRC}}$ RBs to communicate with each other as shown by the blue lines in Figure \ref{Fig:Links}(a). For each AV, it is assigned with $\lfloor \frac{K_{\rm DSRC}}{n_{\textmd{\tiny AV}}} \rfloor$ RBs. Similar to the V2I link, the achievable information rate between the $i$-th VN and the $j$-th VN is
\begin{equation}\label{Eq:CV2V}
C_{j,i}(t) = \bigg\lfloor \frac{K_{\rm{DSRC}}}{n_{\textmd{\tiny AV}}} \bigg\rfloor {\rm log}_{2}\bigg(1+\frac{P_i}{\sigma_V^2 10^{\frac{L_{j,i}(d_{j,i}(t))}{10}}}\bigg),
\end{equation}
where $L_{i,j}$, $d_{i,j}(t)$, $P_i$ and $\sigma_V^2$ are the path-loss, the distance between the $i$th VN and the $j$-th VN at time $t$,  the transmit power of the RV and the AWGN noise power over the DSRC link of each RB, respectively.

\subsubsection{Two-hop relaying Communications} As described above, AV's data is forwarded by RV via out-of-band relaying DSRC V2V links.
Considering the limitations of process capability and the available transmit power of the nodes and also in order to avoid high signaling
overhead and scheduling complexity, we assume that each RV only can help one AV and each AV is only served by one RV at a time. In this case, once a RV is allocated to a AV, the AV will give its assigned V2I RBs to its helping RV and the RV thus can use these RBs to help the information forwarding to AV. This means that, RV does not need to spend its own V2I resources to help the information forwarding, as it has been assigned with
extra V2I resources for relaying.  Therefore, for each RV, its own communication quality with the BS will not be affected when it relays the information for its helped AV.
DF relaying protocol is employed applied at the RV.
Thus, the end-to-end achievable information rate from the BS to the AV $j$ via helping RV $i$ is bounded by the minimal achievable rate of the two hops, which can be expressed by
\begin{equation}\label{Eq:V2I+V2V}
C_{j,i,B}(t) = {\rm min}\{C_{j,i}(t), C_{i,\rm{B}}(t)\}.
\end{equation}

\section{Problem Formulation}\label{Sec:SectIII}
Based on the vehicular network system model described in Section \ref{Sec:SectII}, in this section, we shall discuss how to design efficient scheduling scheme for it.  In order to clearly figure out the
advantages of our proposed MSRS, we shall first explain the shortcomings of existing IRRS method.

For a downlink vehicular network as shown in Figure \ref{Fig:model}, its potential connectivity can be illustrated as shown in Figure \ref{Fig:Links} (a), which looks like a complete graph composed of uni-directional V2I links and bi-directional V2V links. The goal of a scheduling scheme is to achieve the optimal system transmission performance, e.g., the maximum throughput, by jointly scheduling the V2I and V2V links to determine a spanning-tree as shown in Fig \ref{Fig:Links} (b).

Let $\bm{s}=\{\mathbb{N}_{\rm{RV}},\mathbb{N}_{\rm{AV}},\mathbb{N}_{\rm{CV}}\}$ be the vehicle division strategy, $\mathbb{N}_{\rm{RV}}$, $\mathbb{N}_{\rm{AV}}$,  $\mathbb{N}_{\rm{CV}}$, $\mathbb{N}_{\rm{VN}}$ denote the set of RVs, AVs, CVs and all VNs respectively. They satisfy that

\begin{equation} \label{eq:Sets}
\left\{ \begin{aligned}
         \mathbb{N}_{\rm{CV}}&\cup \mathbb{N}_{\rm{RV}} \cup \mathbb{N}_{\rm{AV}} =\mathbb{N}_{\rm{VN}}, \\
         \mathbb{N}_{\rm{CV}}&\cap \mathbb{N}_{\rm{RV}}=\emptyset,\\
         \mathbb{N}_{\rm{RV}}&\cap \mathbb{N}_{\rm{AV}}=\emptyset,\\
         \mathbb{N}_{\rm{CV}}&\cap \mathbb{N}_{\rm{AV}}=\emptyset.\\
\end{aligned} \right.
\end{equation}
Let $n_{\textmd{\tiny RV}}$ and $n_{\textmd{\tiny CV}}$ be the number of vehicles in the set of $\mathbb{N}_{\rm{RV}}$ and
$\mathbb{N}_{\rm{CV}}$. We have that $n_{\textmd{\tiny AV}}+n_{\textmd{\tiny RV}}+n_{\textmd{\tiny CV}}=N$. Moreover, because of the relaying constraint, i.e., each relay only serving one AV, thus it can be inferred that
\begin{equation}\label{eq:NN}
n_{\textmd{\tiny RV}}=n_{\textmd{\tiny AV}}\,\,\textrm{and}\,\,n_{\textmd{\tiny AV}}\leq \frac{N}{2}.
\end{equation}

Let $\bm{\theta}=\{\theta_{i,j}\}_{n_{\textmd{\tiny RV}} \times n_{\textmd{\tiny AV}}}$ be the pairing matrix composed by binary element $\theta_{i,j}\in \{0,1\}$
among the RVs and AVs. Specifically, if RV $i$ is paired with AV $j$, $\theta_{i,j}=1$. Otherwise, $\theta_{i,j}=0$.  Since each RV can only be paired with one AV and vise versa, we have that

\begin{equation} \label{eq:theta}
\left\{ \begin{aligned}
         \sum\nolimits_{i=1}^{N_{\tiny \rm RV}} {\theta _{j,i}}  = 1, \forall j=1,...,n_{\textmd{\tiny AV}},\\
           \sum\nolimits_{j=1}^{n_{\textmd{\tiny AV}}} {\theta _{j,i}}  = 1, \forall i=1,...,n_{\textmd{\tiny RV}}.\\
\end{aligned} \right.
\end{equation}

If current IRRS is adopted, to maximize the total sum-rate of the system, the system resource is scheduled for each time period $T$ between $[t,t+T]$ by solving the following optimization problem.

\begin{flalign}
\max\limits_{\bm{s,\theta}}\,\,&\sum\nolimits_i^{n_{\textmd{\tiny RV}}} {C_{i,\rm{B}}(t)} + \sum\nolimits_j^{n_{\textmd{\tiny AV}}} \theta_{j,i}{C_{j,i,\rm{B}}(t)}  + \sum\nolimits_{\ell}^{n_{\textmd{\tiny CV}}} {C_{\ell,\rm{B}}(t)}\nonumber\\
\textrm{s.t.}\,\,&(\ref{eq:Sets}),(\ref{eq:NN}),(\ref{eq:theta}).
\end{flalign}

\begin{figure}
\centering
\includegraphics[width=0.49\textwidth]{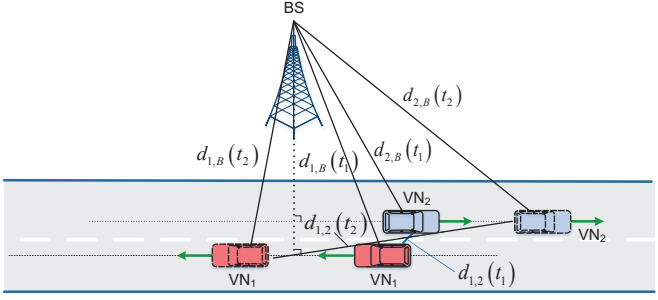}
\caption{The impact of vehicle's movement on system performance}\label{Fig:ImpactVNmovement}
\end{figure}

However, IRRS did not fully consider the impact of mobility on vehicular networks. So it cannot exploit the potential achievable transmission capacity, especially for high mobility scenarios. The reason is that, when the vehicle moves very swiftly, the distance between any two communication nodes in the system will change dramatically along with the time within $T$. Thus, the instantaneous achievable rate calculated at $t$, i.e., $C(t)$, may have an non-negligible deviation compared with that of $t'\in (t,t+T]$, i.e., $C(t')$. This may degrade system performance. For example, as show in Figure \ref{Fig:ImpactVNmovement}, from time $t_1$ to $t_2$, VN$_1$ firstly moves closer to the BS and then passes by the BS and then moves far away from the BS, and VN$_2$ moves far way from the BS. At $t_1$ the distance between the BS and VN$_1$ is $d_{1,{\rm B}}(t_1)$ and that between the BS and VN$_2$ is $d_{2,{\rm B}}(t_1)$, which change to be $d_{1,{\rm B}}(t_2)$ and $d_{1,{\rm B}}(t_2)$. Meanwhile, due to the mobility, the distance between the VNs also changes. The distance between VN$_1$ and VN$_2$ increase from $d_{1,2}(t_1)$ to $d_{1,2}(t_2)$.  Suppose at time $t_1$, VN$_1$ is determined to communication with BS via the help of  VN$_2$. At $t_2$, this choice may also be not efficient any more because  VN$_2$ is also too far way from VN$_1$. Thus, VN$_1$ should select another vehicle to help it. Moreover, suppose at time $t_1$, VN$_2$ is determined to communication with BS via the direct LET-Advanced link. At $t_2$, this may be not efficient any more, due to the long distance between BS and VN$_2$.

If one desires to release the impact of mobility and reduce the performance loss by IRRS, the time period $T$ have to be shortened, which however, will result in the increase of signaling overhead and computational overhead. As a result, the system performance may be deteriorated. In order to overcome the shortcomings of IRRS and make the scheduling be capable of adapting to high mobility case, we introduce the mobile service and then present our proposed MSRS.

\subsection{Mobile Service}
The concept of mobile service can be traced back to \cite{R:CHUANG}, which was proposed to efficiently describe the information transmission capacity of high-speed railway communication system. Here, we introduce it to the heterogenous V2I and V2V coexisting vehicular system.

Due to the channel variation and topology change of vehicular networks, the scheduling should be performed periodically. For the $k-$th period,  the mobile service is defined as
\begin{equation}\label{Eq:service}
S(T_k) = \int_{t_k}^{t_k+T} C(t)dt,
\end{equation}
where $T_k$ represents the $k$-th scheduling period, $t_k$ is the beginning time instant of $T_k$, and $T$ is the time interval of each scheduling period.

It can be seen that the mobile service is an integral of the instantaneous
achievable rate over a given scheduling period $T$, which actually upper bounds the transmission capacity for any given pair of communication nodes, and it represents the maximum amount of data that the physical layer can provide for the network layer.

Substituting (\ref{Eq:capacity})  into (\ref{Eq:service}), we get
\begin{equation}\label{Eq:service2}
S(T_k) = \int_{t_k}^{t_k+T} {\rm log}_{2}\bigg(1+\frac{P_s}{\sigma_0^2 10^{\frac{L(d(t))}{10}}}\bigg)dt.
\end{equation}

Parameters in (\ref{Eq:service2}) are knowable at time $t_k$ except relative distance $d(t)$.  Fortunately, since the movement of the vehicle is limited by the lane and the moving speed of the vehicle can be regarded as constant within a short $T$. Once the initial position of the vehicle is known at $t_k$, which can be realized by using the Global Positioning System (GPS), the instantaneous position of the vehicle for any $t\in(t_k,t_k+T)$ can be calculated. The coordinate of the BS is $(x_{\rm B},y_{\rm B})$ and the position of the $i-$th vehicle is $(x_{i,t_k},y_{i,t_k})$. $v_i$ is the moving speed of the vehicle and $\phi_i$ is the angle between $v_i$ and the $x$-axis of the reference coordinate system. At $t\in(t_k,t_k+T)$, the new position of the vehicle can be predicted by $(x_{i,t} ,y_{i,t})=(x_{i,t_k}+v_i t\cos\phi_i ,y_{i,t_k}+v_it\sin\phi_i)$.
Therefore, at $t$, the distance between the $i-$th VN and the BS is
\begin{flalign}\label{Eq:dib}
&d_{i,\rm{B}}(t)=\\
&\sqrt{(y_{i,t_k}+v_it\sin\phi_i-y_{B})^2+(x_{i,t_k}+v_it\cos\phi_i-x_{B})^2}.\nonumber
\end{flalign}
and the distance between the $i-$th VN and the $j-$th VN is
\begin{flalign}\label{Eq:dij}
d_{j,i}(t)=\sqrt{(y_{i,t}-y_{j,t})^2+(x_{i,t}-x_{j,t})^2}.\nonumber
\end{flalign}
As a result, the V2I mobile service of the $i-$th VN in $T_k$ can be given by
\begin{equation}\label{Eq:V2Iservice2}
S_{i,\rm{B}}(T_k) = \int_{t_k}^{t_k+T}  \bigg\lfloor \frac{K_{\rm{LTE}}}{N} \bigg\rfloor {\rm log}_{2}\bigg(1+\frac{P_{\rm{B}}}{\sigma_I^2 10^{\frac{L_{i,\rm{B}}(d_{i,\rm{B}}(t))}{10}}}\bigg) dt
\end{equation}
and the V2V mobile service of between the $i-$th VN and the $j-$th VN in $T_k$ can be given by
\begin{equation}\label{Eq:V2V}
S_{j,i}(T_k) = \int_{t_k}^{t_k+T} \bigg\lfloor \frac{K_{\rm{DSRC}}}{n_{\textmd{\tiny AV}}} \bigg\rfloor {\rm log}_{2}\bigg(1+\frac{P_i}{\sigma_V^2 10^{\frac{L_{j,i}(d_{j,i}(t))}{10}}}\bigg) dt.
\end{equation}
Similar to (\ref{Eq:V2I+V2V}), the mobile service of the two-hop communication from BS to the $j-$th VN via $i-$th VN also is bounded by the minimal mobile service amount of the two hops. Thus, it can be given by
\begin{equation}\label{Eq:V2I+V2V-S}
S_{j,i,B}(T_k) = {\rm min}\{S_{j,i}(T_k), S_{i,\rm{B}}(T_k)\}.
\end{equation}

\subsection{Maximization Optimization Problem for MSRS}
The goal of our scheduling scheme is to maximize the total mobile service amount by jointly optimizing $\bm{s}$ and $\bm{\theta}$. Hence our optimization problem can be given by

\begin{small}
\begin{flalign}\label{Opt:Ss}
\max\limits_{\bm{s,\theta}}\,\,&\sum\limits_i^{n_{\textmd{\tiny RV}}} {S_{i,\rm{B}}(T_k)} + \sum\limits_j^{n_{\textmd{\tiny AV}}} \theta_{j,i}{S_{j,i,\rm{B}}(T_k)}  + \sum\limits_{\ell}^{n_{\textmd{\tiny CV}}} {S_{\ell,\rm{B}}(T_k)}\\
\textrm{s.t.}\,\,&(\ref{eq:Sets}),(\ref{eq:NN}),(\ref{eq:theta}).\nonumber
\end{flalign}
\end{small}

It can be seen that Problem (\ref{Opt:Ss}) actually maximize the total information transmission amount over a time period $T$. For a fixed $T$, Problem (\ref{Opt:Ss}) equals to maximize the average system throughput in terms of (bit/s).

Note that, to solve Problem (\ref{Opt:Ss}) is not trivial, since it is a combinational optimization problem with very high computational complexity if it is solved in an intuitively way as follows.
\begin{itemize}
  \item  1). Select $n_{\textmd{\tiny AV}}$ VNs from the $N$ VNs as AVs to receive data via two-hop communications. (The
            amount of possible selections is $C_{N}^{n_{\textmd{\tiny AV}}}$, where $C_y^x$ denotes the $x$-combinations over $y$.)
  \item  2). Select $n_{\textmd{\tiny RV}}$ VNs from the rest $N-n_{\textmd{\tiny AV}}$ VNs as RVs to help the
$n_{\textmd{\tiny AV}}$ AVs. (The number of possible choices is $C_{N-n_{\textmd{\tiny AV}}}^{n_{\textmd{\tiny RV}}}$.)
  \item  3). Pair the selected $n_{\textmd{\tiny AV}}$ AVs with the selected $n_{\textmd{\tiny RV}}$ RVs in Step 1 and 2 with a one-to-one manner. (The number of possible  pairings is $P_{n_{\textmd{\tiny RV}}}^{n_{\textmd{\tiny AV}}}$, where $A_y^x$ is the $x$-permutations of $y$ operation.)
       \item 4). Perform Step 1 to 3 above for  $n_{\textmd{\tiny AV}}=1,...,\left\lfloor {\frac{N}{2}} \right\rfloor$ , where $\left\lfloor {\frac{N}{2}} \right\rfloor$ is the integer floor value of $\frac{N}{2}$. (It can be inferred that, $0 \leq n_{\textmd{\tiny AV}}\leq N/2$, as the $n_{\textmd{\tiny AV}}=n_{\textmd{\tiny RV}}$ and $n_{\textmd{\tiny AV}}+n_{\textmd{\tiny RV}}+n_{\textmd{\tiny CV}}=N$.) Then, select the one with maximum total system service amount as the optimal scheduling.
\end{itemize}

It can be observed that for each $n_{\textmd{\tiny AV}}$, the computational complexity from Step 1 to 3 is $C_{N}^{n_{\textmd{\tiny AV}}}C_{N-n_{\textmd{\tiny AV}}}^{n_{\textmd{\tiny RV}}}P_{n_{\textmd{\tiny RV}}}^{n_{\textmd{\tiny AV}}}$. Thus, total computational complexity of the method above is the summation over all $n_{\textmd{\tiny AV}}=0,...,\left\lfloor {\frac{N}{2}} \right\rfloor$, i.e.,
\begin{equation}\nonumber
\sum_{n_{\textmd{\tiny AV}}=0}^{\lfloor N/2 \rfloor}
C_{N}^{n_{\textmd{\tiny AV}}}C_{N-n_{\textmd{\tiny AV}}}^{n_{\textmd{\tiny RV}}}P_{n_{\textmd{\tiny RV}}}^{n_{\textmd{\tiny AV}}}=\sum_{n_{\textmd{\tiny AV}}=0}^{\lfloor N/2 \rfloor}\frac{{N!}}{{n_{\textmd{\tiny AV}}!(N - 2n_{\textmd{\tiny AV}})!}},
\end{equation}
which is extremely complex. Especially, for a relatively large $N$ scenario, this method is unacceptable. Thus, we shall find other way to solve this problem with low complexity.

\section{Low-complex Algorithm Design for MSRS}\label{Sec:SectVI}

\subsection{Problem Analysis and Transfrom}

By observation, we can obtain some disciplines for the optimal result associated with Problem (\ref{Opt:Ss}).

\textbf{Lemma 1.} The optimal solution $\{\bm{s^*},\bm{\theta^*}\}$ of Problem (\ref{Opt:Ss}) satisfies that, for a AV $j\in \mathbb{N}_{\rm{AV}}^*$ and a RV $i\in \mathbb{N}_{\rm{RV}}^*$, if $\theta_{i,j}=1$, the mobile service amount of the AV over the V2I link must be less than that of the RV over the V2I link, i.e., $S_{j,{\rm B}}(T_k)\leq S_{i,{\rm B}}(T_k)$.
\begin{proof}
Suppose $S_{j,{\rm B}}(T_k)> S_{i,{\rm B}}(T_k)$. Denote $\{\bm{s^\sharp},\bm{\theta^\sharp}\}$ be another scheduling which is obtained by picking $x$ out from $\mathbb{N}_{\rm{AV}}^*$ and then putting $x$ into $\mathbb{N}_{\rm{CV}}^*$. It can be easily inferred that the total mobile service amount associated with $\{\bm{s^*},\bm{\theta^*}\}$ is less than that associated with $\{\bm{s^\sharp},\bm{\theta^\sharp}\}$. This contradicts that $\{\bm{s^*},\bm{\theta^*}\}$ is the optimal solution. Therefore, Lemma 1 is proved.
\end{proof}

Lemma 1 indicates that, the VN with smaller V2I link mobile service amount should be selected as AV compared with the one with larger V2I link mobile service amount. Moreover, the CV can be considered as a special RV, which has zero assisted AV. Thus, we can extend $\bm{\theta}$ to be $\bm{\theta}=\{\theta_{i,j}\}_{(N-n_{\textmd{\tiny AV}}) \times n_{\textmd{\tiny AV}}}$ and the constraint (\ref{eq:theta}) is refined as

\begin{equation} \label{eq:theta2}
\left\{ \begin{aligned}
         \sum\nolimits_{i=1}^{1-n_{\textmd{\tiny AV}}} {\theta _{j,i}}= 1, \,\,\,\,\,\,\,\,\,\,\,\,\forall j=1,...,n_{\textmd{\tiny AV}},\\
           \sum\nolimits_{j=1}^{n_{\textmd{\tiny AV}}} {\theta _{j,i}}  \leq 1, \,\,\forall i=1,...,N-n_{\textmd{\tiny AV}}.\\
\end{aligned} \right.
\end{equation}
and $\bm{s}$ is refined to be $\bm{s}=\{\mathbb{N}_{\rm{AV}},\mathbb{N}_{\rm{VN}}-\mathbb{N}_{\rm{AV}}\}$. By doing so, the variables $\mathbb{N}_{\rm{CV}}$ and $\mathbb{N}_{\rm{RV}}$ are merged into (\ref{eq:theta2}) and  Problem (\ref{Opt:Ss}) can be equivalently rewritten by a simpler expression as

\begin{flalign}\label{Opt:Ssn}
\max\limits_{\bm{s,\theta}}\,\,&\sum\nolimits_i^{N-n_{\textmd{\tiny AV}}} {S_{i,\rm{B}}(T_k)} + \sum\nolimits_j^{n_{\textmd{\tiny AV}}} {S_{j,i,\rm{B}}(T_k)}\\
\textrm{s.t.}\,\,&(\ref{eq:theta2}).\nonumber
\end{flalign}

\subsection{Our Proposed Algorithm}
Based on these observations, we design an efficient scheduling algorithm with low-complexity, which is composed of the following four steps.
%
%\begin{algorithm}[htb]
%\caption{ Basic framework of our proposed low-complexity algorithm.}
%\label{alg:ALg1}
%\textbf{Step 1.} Calculate the V2I mobile service amount $S_{n,{\rm B}}$ for all VNs in the system, i.e., $n=1,..., N$;\\
%\textbf{Step 2.} Sort the $N$ VNs in  an descending order in terms of the V2I service amount $S_{n,{\rm B}}$;\\
%\textbf{Step 3.} For a given $n_{\textmd{\tiny AV}}$, select the last $n_{\textmd{\tiny AV}}$ VNs as AVs and the rest $N-n_{\textmd{\tiny AV}}$ VNs as RVs, and then find the optimal $\bm{\theta}^*$ under constraint (\ref{eq:theta2}) in terms of Algorithm \ref{alg:ALg2} and calculate the corresponding system total mobile service;\\
%\textbf{Step 4.} Update $n_{\textmd{\tiny AV}}$ and repeat Step 3. Until the system total mobile service can not be increased any more, Algorithm \ref{alg:ALg1} is stopped. Then the optimal $n_{\textmd{\tiny AV}}^*$ and $\bm{\theta}^*$ is obtained.
%\end{algorithm}
%
%The detailed operation of each step of Algorithm \ref{alg:ALg1} is presented as follows.

\subsubsection{\textsf{\textbf{Step 1}}. Calculate the V2I Mobile Service Amount}
The V2I mobile service amount $S_{n,{\rm B}}$ for all $n=1,..., N$ can be directly calculated in terms of (\ref{Eq:V2Iservice2}).

\subsubsection{\textsf{\textbf{Step 2}}. Sort all VNs in terms of V2I Mobile Service Amount} The VNs can be sorted in a descending by some typical low-complex sorting methods such as the bubble sorting and the selection sorting. For convenience we denote the set of the sorted VNs as $\mathbb{N}_{\rm{VN}}'$.

\subsubsection{\textsf{\textbf{Step 3}}. Find the optimal $\bm \theta^*$ for a given $n_{\textmd{\tiny AV}}$}
For a given $n_{\textmd{\tiny AV}}$, we select the last $n_{\textmd{\tiny AV}}$ VNs in $\mathbb{N}_{\rm{VN}}'$ as AVs and put them in $\mathbb{N}_{\rm{AV}}$, and denote the set composed of the rest $N-n_{\textmd{\tiny AV}}$ VNs as $\mathbb{N}_{\rm{temp}}$. This means ${\bm s}=\{\mathbb{N}_{\rm{AV}},\mathbb{N}_{\rm{VN}}-\mathbb{N}_{\rm{AV}}\}$. In this case, Problem (\ref{Opt:Ssn}) is reduced to be
\begin{flalign}\label{Opt:Sst}
\max\limits_{\bm{\theta}}\,\,&\sum\nolimits_i^{N-n_{\textmd{\tiny AV}}} {S_{i,\rm{B}}(T_k)} + \sum\nolimits_j^{n_{\textmd{\tiny AV}}} \theta_{j,i}{S_{j,i,\rm{B}}(T_k)}\\
\textrm{s.t.}\,\,&(\ref{eq:theta2}).\nonumber
\end{flalign}

Let $n_{\textmd{\tiny RV}}'=N-n_{\textmd{\tiny AV}}$. Define a matrix \textbf{W}, $\textbf{W}^T=\{S_{j,i,{\rm B}}\}_{n_{\textmd{\tiny AV}}\times n_{\textmd{\tiny RV}}'}$, i.e.,
\begin{equation}\label{Eq:w-matrix1}
\textbf{W}=\left( {\begin{array}{*{20}{c}}
S_{1,1,{\rm B}}&S_{1,2,{\rm B}}& \cdots &S_{1,n_{\textmd{\tiny AV}},{\rm B}}\\
S_{2,1,{\rm B}}&S_{2,2,{\rm B}}& \cdots &S_{2,n_{\textmd{\tiny AV}},{\rm B}}\\
 \vdots & \vdots & \ddots & \vdots \\
S_{n_{\textmd{\tiny RV}}',1,{\rm B}}&S_{n_{\textmd{\tiny RV}}',2,{\rm B}}& \cdots &S_{n_{\textmd{\tiny RV}}',n_{\textmd{\tiny AV}},{\rm B}}\\
\end{array}}\right)
\end{equation}
which can be constructed in terms of (\ref{Eq:V2I+V2V-S}).

The matrix $\textbf{W}$ can be regarded as a benefit matrix, whose
column indices can be regarded as different operators and
row indices can be regarded as different machines, i.e., a
total of $N-n_{\textmd{\tiny AV}}$ different machines to be operated by $n_{\textmd{\tiny AV}}$ different
operators. Thus, the value of each entry can be considered as
the benefit from operating a specific machine by a specific
operator. Then the problem in (\ref{Opt:Sst}) is equivalent to maximizing
the sum benefit via choosing an exactly one element in each
row and each column of $\textbf{W}$, where each operator is only
permitted to operate only one machine. This is a
assignment problem.

\textbf{Case I}: $n_{\textmd{\tiny AV}}=n_{\textmd{\tiny RV}}'$. When $n_{\textmd{\tiny AV}}=n_{\textmd{\tiny RV}}'=\frac{N}{2}$, $\textbf{W}$ is square matrix. The assignment problem can be solved by
the Hungarian algorithm \cite{R:Khun}, whose complexity is
$\mathcal{O}(N^3)$. For emphasis, the algorithm is described as Algorithm \ref{alg:ALg2}.

\begin{algorithm}[htb]
\caption{Calculating the optimal pairing over an input square $\textbf{W}$.}
\label{alg:ALg2}
\textbf{a)} Modify all elements in $\textbf{W}$ in terms of $\max\limits_{1\leq i,j\leq n_{\textmd{\tiny RV}}'}\{S_{j,i,{\rm B}}\}-S_{j,i,{\rm B}}$ for all $1\leq i,j\leq n_{\textmd{\tiny RV}}'$;\\
\textbf{b)} Subtract the elements in each row from the maximum
number in the row, and subtract the minimum number in each
column from the whole column of $\textbf{W}$;\\
\textbf{c)} Cover all zero-elements in $\textbf{W}$ as few lines as possible;\\
\textbf{d)} If the number of lines equal to the size of $\textbf{W}$, the optimal
solution is found. Otherwise, find the minimum number that is
uncovered. Subtract this minimum number from all uncovered
elements and add it into values at the intersections of lines,
then go to Step b).
\end{algorithm}

\textsf{Case II}: $n_{\textmd{\tiny AV}}\neq n_{\textmd{\tiny RV}}'$. When $n_{\textmd{\tiny AV}}\neq n_{\textmd{\tiny RV}}'$, according to (\ref{eq:NN}), it can be inferred that $n_{\textmd{\tiny AV}}< n_{\textmd{\tiny RV}}'$. In this case, the Algorithm \ref{alg:ALg2} can not be directly adopted because $\textbf{W}$ is a non-square matrix. Thus, we firstly extend $\textbf{W}$ to a square matrix $\widetilde{\textbf{W}}$ by adding $n_{\textmd{\tiny RV}}'-n_{\textmd{\tiny AV}}$ columns with zero-elements as (\ref{Eq:w-matrix2}).

\begin{flalign}\label{Eq:w-matrix2}
&\widetilde{\textbf{W}}=[\textbf{W}_{n_{\textmd{\tiny RV}}'\times n_{\textmd{\tiny AV}}}\,\,\textbf{0}_{n_{\textmd{\tiny RV}}'\times (n_{\textmd{\tiny RV}}'-n_{\textmd{\tiny AV}})}]=\\
&\left( {\begin{array}{*{20}{c}}
S_{1,1,{\rm B}}&S_{1,2,{\rm B}}& \cdots &S_{1,n_{\textmd{\tiny AV}},{\rm B}}&0&\cdots&0\\
S_{2,1,{\rm B}}&S_{2,2,{\rm B}}& \cdots &S_{2,n_{\textmd{\tiny AV}},{\rm B}}&0&\cdots&0\\
 \vdots & \vdots & \ddots & \vdots&\vdots&\ddots&\vdots\\
S_{n_{\textmd{\tiny RV}}',1,{\rm B}}&S_{n_{\textmd{\tiny RV}}',2,{\rm B}}& \cdots &S_{n_{\textmd{\tiny RV}}',n_{\textmd{\tiny AV}},{\rm B}}&0&\cdots&0\nonumber
\end{array}}\right)
\end{flalign}

Then, Algorithm \ref{alg:ALg2} can be adopted to calculate the optimal pairing over the extended square matrix $\widetilde{\textbf{W}}$.

%1) Add $n_{\textmd{\tiny RV}}'-n_{\textmd{\tiny AV}}$ virtual AVs,
%
%The mobile service of the $N_R$ RVs is fixed after the classifying, so the key to maximize the throughput and spectral efficiency of vehicular networks is to ensure each NV pair off
%with an appropriate RV. This problem can be formulated as maximum weighted matching (MWM) in a weighted bipartite graph. We construct the weighted bipartite graph as following.
%\begin{itemize}
%  \item  The RVs and AVs constitute the two mutually disjoint vertex set $U_R$ and $U_D$ respectively.
%  \item  The $N_R*N_D$ links between the RVs in $U_R$ and the AVs in $U_D$ constitute the edge set $E$, $e_{ij}\in E$ and $i\in U_R, j\in U_D$.
%  \item  The weight of edge $e_{ij}$ is $w(e_{ij})=S_{j,i,B}(T_k)$.
%\end{itemize}
The matrix extension process can be explained by the following example. Given a non-square $5\times 4$ matrix $\textbf{W}$ as
\begin{equation}\label{Eq:w-exp}
\begin{array}{l}
\begin{array}{*{20}{c}}
\,\,\quad\quad\quad&\quad&\textrm{AV}_1&\textrm{AV}_2&\textrm{AV}_3&\textrm{AV}_4
\end{array}\\
\textbf{W}=
\begin{array}{*{20}{c}}
\textrm{RV}_1\\
\textrm{RV}_2\\
\begin{array}{l}
\textrm{RV}_3\\
\textrm{RV}_4\\
\textrm{RV}_5
\end{array}
\end{array}\left( {\begin{array}{*{20}{c}}
\,2\quad&3\quad&\,0\quad&\,1\\
\,3\quad&2\quad&\,3\quad&\,6\\
\,4\quad&0\quad&\,3\quad&\,0\\
\,5\quad&2\quad&\,4\quad&\,6\\
\,1\quad&0\quad&\,0\quad&\,2\\
\end{array}} \right),
\end{array}
\end{equation}
whose row index and column index represents the RV and AV, respectively. Since there are 5 RVs and 4 AVs, one zero-column is added. Thus, the extended square matrix $\widetilde{\textbf{W}}$ is
\begin{equation}\label{Eq:w-exp2}
\begin{array}{l}
\begin{array}{*{20}{c}}
\,\,\quad\quad\quad&\quad&\textrm{AV}_1&\textrm{AV}_2&\textrm{AV}_3&\textrm{AV}_4& {\underline{\textrm{AV}_5}}
\end{array}\\
\widetilde{\textbf{W}}=
\begin{array}{*{20}{c}}
\textrm{RV}_1\\
\textrm{RV}_2\\
\begin{array}{l}
\textrm{RV}_3\\
\textrm{RV}_4\\
\textrm{RV}_5
\end{array}
\end{array}\left( {\begin{array}{*{20}{c}}
\,2\quad&3\quad&\,0\quad&\,1\quad&\,0\\
\,3\quad&2\quad&\,3\quad&\,6\quad&\,0\\
\,4\quad&0\quad&\,3\quad&\,0\quad&\,0\\
\,5\quad&2\quad&\,4\quad&\,6\quad&\,0\\
\,1\quad&0\quad&\,0\quad&\,2\quad&\,0\\
\end{array}} \right).
\end{array}
\end{equation}
%The bipartite graph of $\widetilde{\textbf{W}}$ is shown in Figure \ref{Fig:BG}(b), where AV$_5$ is the newly added vertex, the weights of whose edges are all zero. AV$_5$ actually can be considered as a virtual AV node. Thus, the operation of extending $\textbf{W}$ is also equal to the operation of adding $n_{\textmd{\tiny RV}}'-n_{\textmd{\tiny AV}}$ AVs with \textit{zero}-weight edges on the bipartite graph of $\textbf{W}$.
Then $\widetilde{\textbf{W}}$ is a $5\times 5$ square matrix. Algorithm \ref{alg:ALg2} can be adopted. After performing Algorithm \ref{alg:ALg2} on $\widetilde{\textbf{W}}$, the result is shown as
\begin{equation}\label{Eq:w-exp3}
\begin{array}{l}
\begin{array}{*{20}{c}}
\quad&\quad&\textrm{AV}_1&\textrm{AV}_2&\textrm{AV}_3&\textrm{AV}_4&{\underline{\textrm{AV}_5}}
\end{array}\\
\begin{array}{*{20}{c}}
\textrm{RV}_1\\
\textrm{RV}_2\\
\begin{array}{l}
\textrm{RV}_3\\
\textrm{RV}_4\\
\textrm{RV}_5
\end{array}
\end{array}\left( {\begin{array}{*{20}{c}}
\,2\quad&{\textbf{\underline 3}}\quad&\,0\quad&\,1\quad&\,0\\
\,3\quad&2\quad&\,3\quad&\,{\textbf{\underline 6}}\quad&\,0\\
\,4\quad&0\quad&\,{\textbf{\underline 3}}\quad&\,0\quad&\,0\\
\,{\textbf{\underline 5}}\quad&2\quad&\,4\quad&\,6\quad&\,0\\
\,1\quad&0\quad&\,0\quad&\,2\quad&\,{\textbf{\underline 0}}\\
\end{array}} \right).
\end{array}
\end{equation}
where the underlined bold elements are selected as the optimal pairing result. The corresponding optimal $\bm{\theta^*}$ is
\begin{equation}\label{Eq:w-exp3}
\bm{\theta^*}=\left( {\begin{array}{*{20}{c}}
{\rm{0}}&{\rm{1}}&{\rm{0}}&{\rm{0}}&{\rm{0}}\\
{\rm{0}}&{\rm{0}}&{\rm{0}}&{\rm{1}}&{\rm{0}}\\
{\rm{0}}&{\rm{0}}&{\rm{1}}&{\rm{0}}&{\rm{0}}\\
{\rm{1}}&{\rm{0}}&{\rm{0}}&{\rm{0}}&{\rm{0}}\\
{\rm{0}}&{\rm{0}}&{\rm{0}}&{\rm{0}}&{\underline {\rm{0}} }
\end{array}} \right),
\end{equation}
which can be obtained by the following rule as
\begin{flalign}\label{Eqn:Theta2}
{\theta_{j,i}^*}=\left\{ \begin{array}{l}
 1,\,\,\,\,\textrm{if} \,\,S_{j,i,B}\,\,\textrm{is selected and} \,\,\textrm{if column}\\\quad\,\,\,i\,\,\textrm{is not newly added column,}\\
 0,\,\,\,\,\rm{otherwise.}
 \end{array} \right.
 \end{flalign}
%\begin{figure}
%\centering
%\includegraphics[width=0.45\textwidth]{fig3.eps}
%\caption{Example of the BG in an 9-VN vehicular network.}\label{Fig:BG}
%\end{figure}

According to $\bm{\theta^*}$, the RVs and CVs can be determined. That is, for the row whose elements are all zero, it is a CV. Otherwise, it is a RV.
\subsubsection{\textsf{\textbf{Step 4}}. Find the optimal $n_{\textmd{\tiny AV}}^*$}
According to the Step 3 above, for a given $n_{\textmd{\tiny AV}}$, the optimal
$\bm{\theta^*}$ can be obtained in terms of (\ref{Eqn:Theta2}) and the corresponding total mobile service amount is
\begin{equation}
S_{\rm tot}(n_{\textmd{\tiny AV}})=\sum\limits_i^{N-n_{\textmd{\tiny AV}}} {S_{i,\rm{B}}(T_k)} + \sum\limits_j^{n_{\textmd{\tiny AV}}} \theta_{j,i}^*{S_{j,i,\rm{B}}(T_k)},
\end{equation}
which means for different values of $n_{\textmd{\tiny AV}}$, different values of $S_{\rm tot}(n_{\textmd{\tiny AV}})$ can be achieved. Thus, to seek the maximum
total mobile service of the network, an appropriate $n_{\textmd{\tiny AV}}$ is required to be found.

\textbf{Remark:} There exists an optimal $n_{\textmd{\tiny AV}}$ for
the proposed algorithm.

Analysis: For a relatively small $n_{\textmd{\tiny AV}}$, only a few vehicles with small mobile service amount are assisted by RVs, which may cause  inefficient utilization of the
out-of-band DSRC radio resources for V2V links. An extreme case is that when $n_{\textmd{\tiny AV}}=0$, no V2V resource is used. For a relatively large $n_{\textmd{\tiny AV}}$, too many AVs require  help from RVs, according to (\ref{Eq:CV2V}) and (\ref{Eq:V2V}), only a few DSRC bandwidth can be assigned to each AV and the service amount will be limited by the achievable mobile service amount of V2V links, which may result in the decline of the total service amount of the system. Therefore, there is a tradeoff between $n_{\textmd{\tiny AV}}$ and $S_{\rm tot}(n_{\textmd{\tiny AV}})$, which means that there exists an optimal $n_{\textmd{\tiny AV}}$ for
our proposed algorithm.

Based on the above analysis, the optimal $n_{\textmd{\tiny AV}}^*$ can be determined by using some fast search methods such as binary section search or golden search.
For example, one can initialize $n_{\textmd{\tiny AV}}=\left\lfloor {\frac{N}{2}} \right\rfloor$ and $n_{\textmd{\tiny AV}}=0$ for the first and the second round of the extenuation of Step 3. Then, $n_{\textmd{\tiny AV}}$ is updated by using the binary section search or golden search. Once the search method stops, $n_{\textmd{\tiny AV}}^*$ is determined.

\subsection{Complexity Analysis of our Proposed Scheme}
The computational complexity of Step 1 of our algorithm is $\mathcal{O}(N^2)$. The worst computational complexity of the sort method in Step 2 of our algorithm is $\mathcal{O}(N^2)$. The computational complexity of the Hungarian algorithm in Step 3 of our algorithm is $\mathcal{O}(N^3)$ and that of the Binary section search or golden search is $\mathcal{O}(\log_2N)$. Thus, it can be inferred that our proposed  scheme is with a polynomial computational complexity of
$\mathcal{O}(N^3\log_2N)$.

\section{Simulation and Numerical Results}\label{Sec:SectIV}
In this section, we present some simulation and numerical
results to discuss the system performance of our proposed MSRS with respect to VN's number and VN's moving speed. For comparison, we also simulate the optimal scheme, IRRS and traditional non-cooperation scheme as benchmark schemes. Specifically, for \textit{optimal scheme}, it finds the maximum mobile service amount via exhaustive computer search. For \textit{IRRS}, it draws up the scheduling according to VN's instantaneous achievable rate at the time $t_k$ and its mobile service amount is accumulated over the time interval $[t_k,t_k+T]$ according to its assigned scheduling result. For \textit{the traditional non-cooperation scheme}, it schedules all $N$ VNs to directly communicate with BS without using of V2V links.

\subsection{Simulation Scenario and Parameter Configurations}
We simulate the scenario as shown in Figure \ref{Fig:model}, where the vehicles straightly move on the high way towards two opposing directions. A  BS is placed on the roadside with the distance of 15m away from the road, whose position is also considered as the reference point. Each vehicle is represented by a vector $(x,y,v)$, where $x$ and $y$ are the coordinates of $x-$axis and $y-$axis, respectively, and $v$ is the moving speed of the vehicle. $(x,y,v)$ is randomly generated with the limitation of the road area and the vehicle's speed setting.
The coverage radius of the base station is set as 0.5km and the vehicle's moving speed is limited within 35m/s. These configurations will not change in the sequel unless otherwise specified.

In the simulations, the V2I communication is performed over  LTE-A links,
which transmits data via a bandwidth of 40 MHz at a frequency
of 2 GHz with transmit power of 52 dBm\cite{R:3GPP}. The V2V
communication is performed over  DSRC links, which transmits data via a bandwidth of 5 MHz at a frequency of 5.9 GHz with transmit power of 20 dBm\cite{R:DSRC}. The path-loss models of V2I and V2V links used
in the simulations are given by \cite{R:3GPP} and \cite{R:DSRC2} as
\begin{flalign}\nonumber
&L_{\rm LTE}(d(t))=128.1 + 37.6 {\rm log}_{10}\big(\tfrac{d(t)}{1000}\big),\nonumber\\
&L_{\rm DSRC}(d(t))=43.9 + 27.5 {\rm log}_{10}d(t),\nonumber
\end{flalign}
respectively.

\subsection{Performance Comparison with the Optimal Results}
\begin{figure}
\centering
\includegraphics[width=0.475\textwidth]{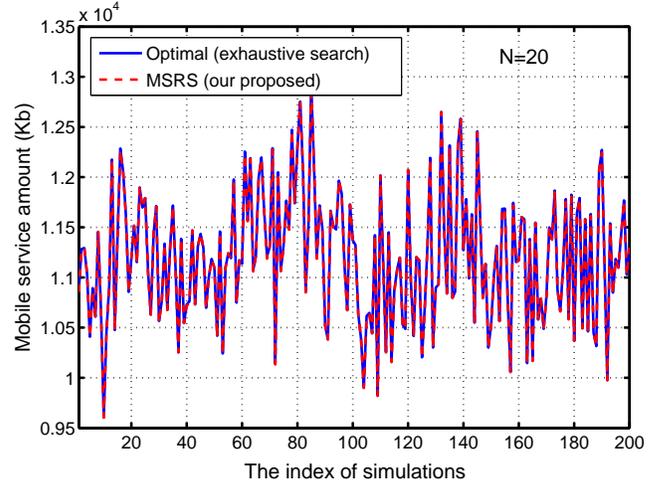}
\caption{Total mobile service amount comparison of MSRS and the exhaustive search for $N=20$.}\label{Fig:Optimal-MSRS0}
\end{figure}

\begin{figure}
\centering
\includegraphics[width=0.475\textwidth]{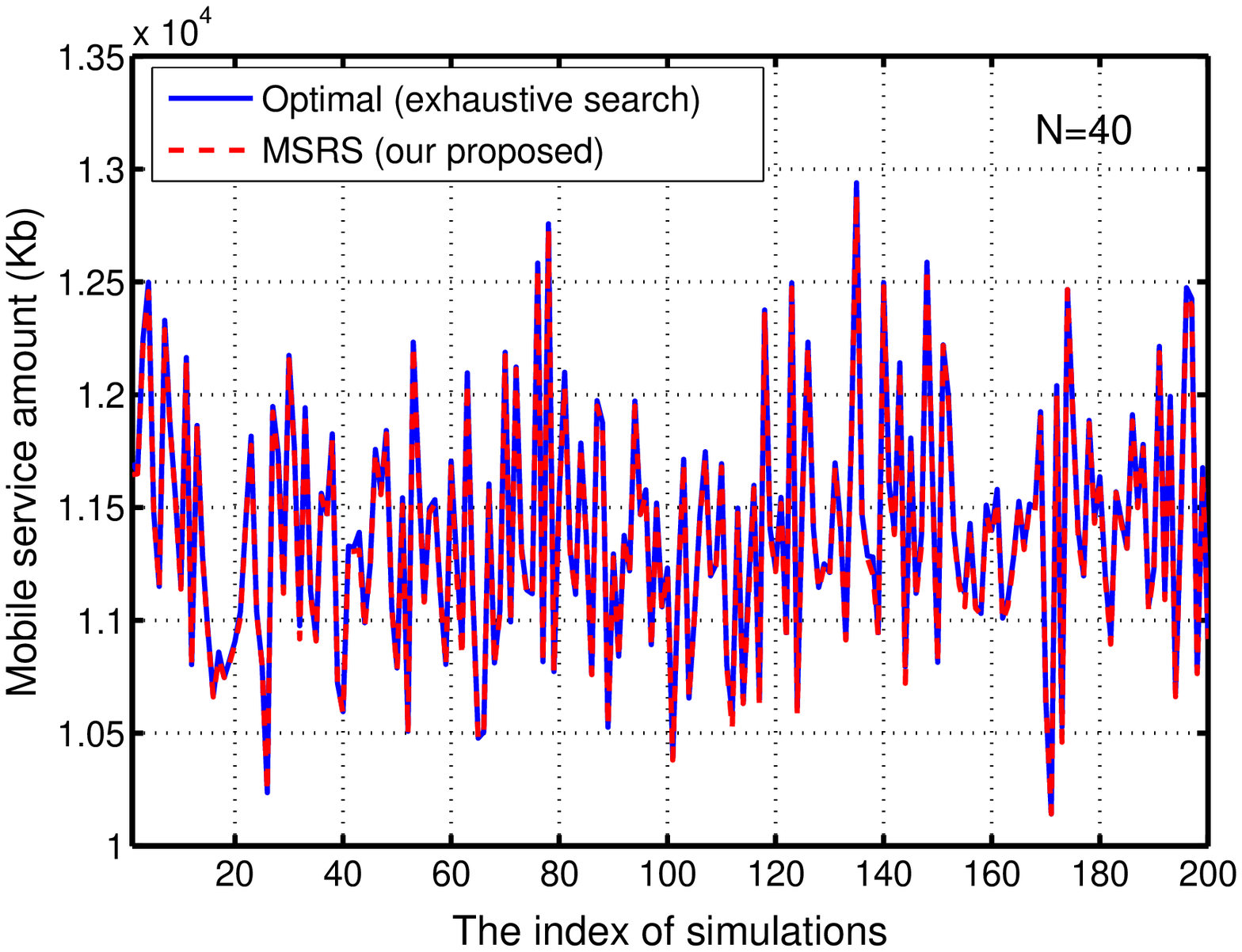}
\caption{Total mobile service amount comparison of MSRS and the exhaustive search for $N=40$.}\label{Fig:Optimal-MSRS1}
\end{figure}

Figure \ref{Fig:Optimal-MSRS0} and Figure \ref{Fig:Optimal-MSRS1} compares the mobile service amount of our proposed MSRS with the optimal results obtained by computer search for $N=20$ and $N=40$, respectively.
In both Figure \ref{Fig:Optimal-MSRS0} and Figure \ref{Fig:Optimal-MSRS1}, the results of  200 simulations are plotted.
 It can be seen that the  mobile service amounts obtained by our proposed MSRS matches the optimal ones very well for both $N=20$ and $N=40$.

\begin{figure}
\centering
\includegraphics[width=0.475\textwidth]{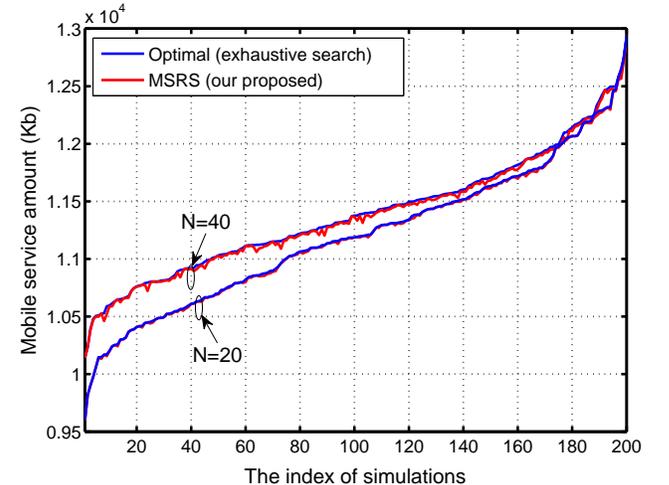}
\caption{Total mobile service amount comparison of MSRS and the exhaustive search in ascending order for $N=20$ and $N=40$.}\label{Fig:Optimal-MSRS2}
\end{figure}

To clearly show the performance difference of the two schemes, we sort the 200 simulations in ascending order in terms of the optimal results. The sorted results of  Figure \ref{Fig:Optimal-MSRS0} and Figure \ref{Fig:Optimal-MSRS1} are presented in Figure \ref{Fig:Optimal-MSRS2}. From Figure \ref{Fig:Optimal-MSRS2}, it can be seen that the performance difference between MSRS and the  exhaustive search is very small, which means that our proposed low-complex MSRS can work well like optimal scheduling.

\begin{figure}
\centering
\includegraphics[width=0.475\textwidth]{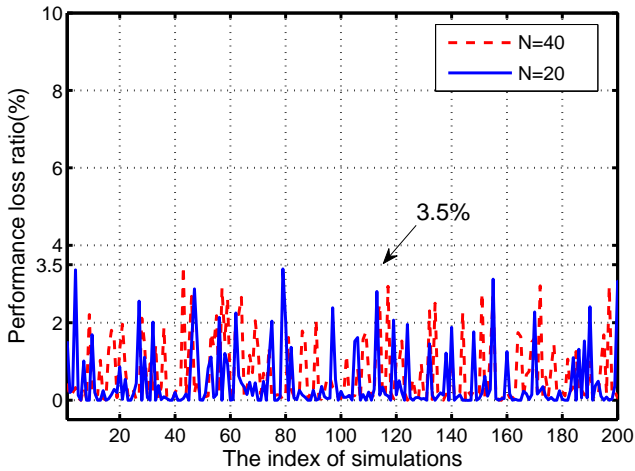}
\caption{Performance loss ratio of MSRS to the exhaustive search for $N=20$ and $N=40$.}\label{Fig:Optimal-MSRS3}
\end{figure}

In  order to further quantify the performance loss of MSRS compared with the optimal result. We define the performance loss ratio as
\begin{equation}\label{Eq:plr}
\textrm{Performance loss ratio (\%)}=\frac{S_{\rm opt}-S_{\rm MSRS}}{S_{\rm opt}},
\end{equation}
where $S_{\rm opt}$ and $S_{\rm MSRS}$ are the mobile service amount obtained by the exhaustive search and MSRS. In terms of (\ref{Eq:plr}) we calculate the performance loss ratio by using 200 simulations in both  Figure \ref{Fig:Optimal-MSRS0} and Figure \ref{Fig:Optimal-MSRS1} and present the results in Figure \ref{Fig:Optimal-MSRS3}.
From Figure \ref{Fig:Optimal-MSRS3}, it can be seen that the maximum performance loss ratio of MSRS to the optimal scheme is less than 3.5\%, which means that MSRS can approximate the optimal results with  at least 96.5\%.

\subsection{Performance Comparison with IRRS and Tractional Non-cooperative Scheme}

\begin{figure}
\centering
\includegraphics[width=0.475\textwidth]{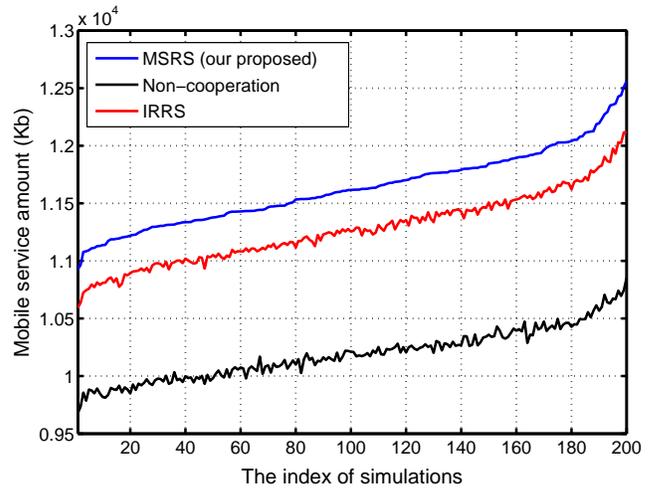}
\caption{The total mobile service amount comparison of MSRS, IRRS and traditional non-cooperation scheme}\label{Fig:MSRS-IRRS-NON}
\end{figure}

Then we compare the mobile service amount of our proposed MSRS with that of IRRS and the traditional non-cooperation scheme. The results is shown in Figure \ref{Fig:MSRS-IRRS-NON}, where 200 simulation results are plotted for $N=100$. Note that, the results in Figure \ref{Fig:MSRS-IRRS-NON} are also the sorted results in ascending order in terms of the mobile service amount of MSRS.

It also can be observed that, the total mobile service amount of MSRS is about 15 percents greater than that of the traditional non-cooperation scheme, which declares that by scheduling the VNs relatively close to the BS as helping RVs to perform cooperative relaying can greatly improve the system mobile service amount, since MSRS brings the cooperative communication benefit to vehicular networks. It also implies that by employing MSRS, the system average throughput and average spectral efficiency of vehicular networks can be greatly enhanced. Moreover, in Figure \ref{Fig:MSRS-IRRS-NON}, it also can be observed that the total mobile service amount of MSRS are about 3.63 percent greater than that of IRRS. The reason is that MSRS is capable of characterizing the channel variations and topology changes of vehicular networks more accurately compared with IRRS.

\subsection{Impacts of System Parameters on Performance}

\subsubsection{The Impact of VN's Number on System Performance}
\begin{figure}
\centering
\includegraphics[width=0.475\textwidth]{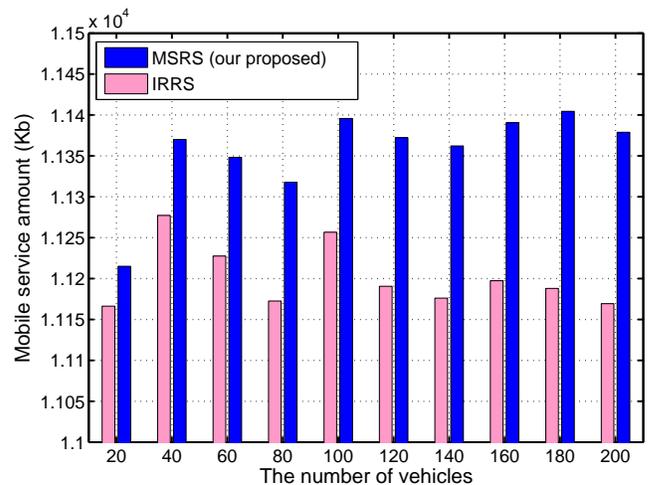}
\caption{The total mobile service amount of MSRS and IRRS versus the number of VNs.}\label{Fig:VNnumber-straight1}
\end{figure}

The mobile service amount of MSRS versus the number of VNs is plotted in Figure \ref{Fig:VNnumber-straight1}, where that of IRRS is also simulated as a benchmark. $N$ is increased from 20, 40 , \ldots, to 200.  For each $N$, the plotted result is obtained by averaging 1000 independent implementations.
It can be seen that, with the increment of the number of VNs, the mobile service amount of MSRS increases basically, but the increasing rate decreases with the increment of $N$. The reason is that more VNs may result in more AVs, which may decease the RBS allocated to each VN and AV. Meanwhile, more VNs may provide more opportunities to find appropriate RV to help the AV. But for IRRS, its mobile service amount basically deceases  with the increment of $N$.

\subsubsection{The Impact of Vehicles' Moving Speed on System Performance}
\begin{figure}
\centering
\includegraphics[width=0.475\textwidth]{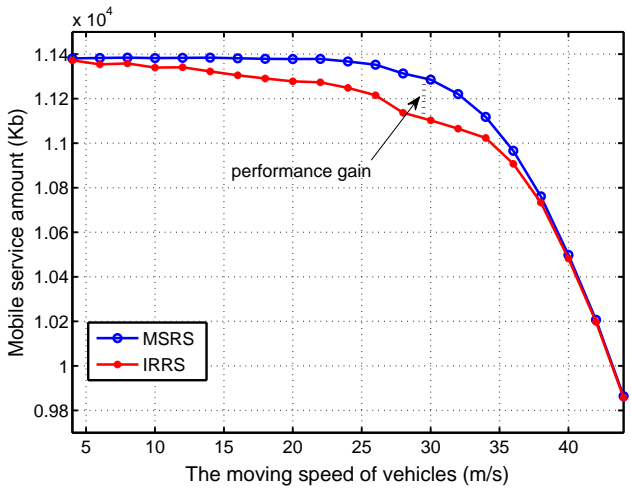}
\caption{The total mobile service amount of MSRS and IRRS versus vehicles' moving speed for $N=50$.}\label{Fig:VNspeed-50}
\end{figure}

\begin{figure}
\centering
\includegraphics[width=0.475\textwidth]{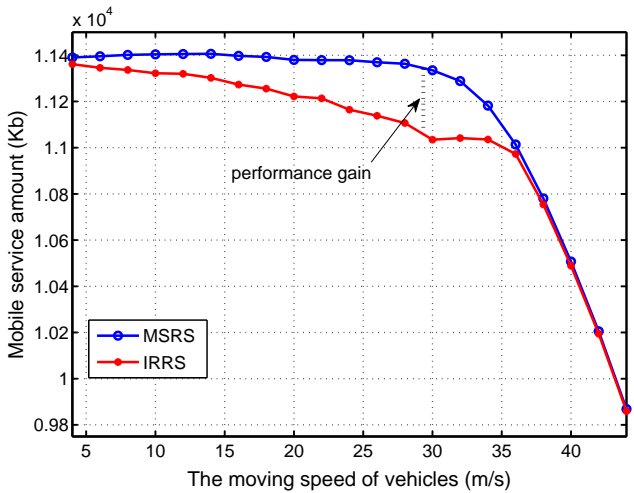}
\caption{The total mobile service amount of MSRS and IRRS versus vehicles' moving speed for $N=100$.}\label{Fig:VNspeed-100}
\end{figure}

The mobile service amounts of both MSRS and IRRS versus vehicles' moving speed are presented in Figure \ref{Fig:VNspeed-50} and Figure \ref{Fig:VNspeed-100} for $N=50$ and $N=100$, respectively. The vehicle's moving speed is increased from 4m/s to 44m/s. The result shown in Figure \ref{Fig:VNspeed-50} and Figure \ref{Fig:VNspeed-100} for each value of speed was also obtained by averaging 1000 simulations.

In the two figures, one can see that the mobile service amount of both MSRS and IRRS decreases with the increment of vehicles' moving speed, but the mobile service amount of both MSRS decrease much slower than that of IRRS. Moreover, as shown in the two figures, when the vehicles are with a relatively high moving speed, e.g., 10m/s to 38m/s, the performance gain of MSRS to IRRS becomes large. When the vehicles are with a low moving speed and very high moving speed, the performance gain of MSRS to IRRS is small.
The reason is that in low mobility case, the inaccuracy of using the achievable instantaneous information rate to describe the system performance and to perform the scheduling is not large, and in very high mobility case, the moving spans of the vehicles within $T$ are too large, which may increase the average distance between any pair of transmit and receiver, consequently decreasing the total mobile service amount of the whole system.

Nevertheless, it can be seen that MSRS always outperforms IRRS for various moving speeds. This demonstrates that MSRS has great potential in enhancing the information transmission capacity for mobile vehicular networks.

%\subsubsection{The Impact of Coverage Radius of BS on System Performance}
%In Figure \ref{Fig:BScoverage-straight1}, the mobile service amount of MSRS and IRRS versus the coverage radius of BS are plotted.
%The coverage radius of BS is increased from 0.5km to 1.5km. The result in each case is obtained by averaging the 1000 implementations.
%It can be seen that the mobile service amounts of both MSRS and IRRS  decreases with the increment of coverage radius of BS. The reason is that for a given $N$, the larger the radius, the sparser the VNs (due to the maximum active VN number and channel resource being fixed) , which may enlarge the distance between BS and the VNs as well as the distance between any two VNs. This may decrease the system information transmission performance.
%
%\begin{figure}
%\centering
%\includegraphics[width=0.475\textwidth]{fig11.eps}
%\caption{The total mobile service amount of MSRS and IRRS versus of BS cell radius}\label{Fig:BScoverage-straight1}
%\end{figure}

\section{Conclusion}\label{Sec:SectV}
In this paper, we introduced the mobile service amount to characterize the information transmission capacity of high-mobility V2V and V2I links. Based on this, we investigated the optimal relaying scheduling scheme to maximize the system total mobile service amount. In order to explore the system information transmission performance limit, we formulated an optimization problem by jointly scheduling the V2I and V2V links. Since the problem is too complex to solve, we designed an efficient algorithm with low-complexity for it. Simulation results demonstrated that our proposed MSRS is able to work well like the optimal scheduling and our proposed MSRS is much more efficient for high-mobility vehicular systems.

%\section*{Acknowledgment}
%This work was supported by  the National Nature Science Foundation of China, no. 61201203, partly by ``973" program, no. 2012CB316100(2), by the State Key Laboratory of Rail Traffic Control and Safety, no. RCS2012ZT008, Beijing Jiaotong University and also by the Fundamental Research Funds for the Central Universities, no. 2014JBM024.

\ifCLASSOPTIONcaptionsoff
  \newpage
\fi

\begin{biography}{Yu Zhang}received the B.S. and Ph.D. degrees from
Beijing Jiaotong University (BJTU), Beijing, China,
in 2004 and 2014, respectively. Since April 2014, he has been a Postdoctoral Research Fellow with the School of Computer and Communi-
cation Engineering, University of Science and Technology Beijing. He has published
more than 30 academic papers in referred journals
and conferences. Dr. Zhang now serves as a TPC Chair of The Third International Conference on Cyberspace Technology 2015. His current research interests include
wireless cooperative networks and signal processing. 

\end{biography}
\begin{biography}{Ke Xiong}received the B.S. and Ph.D. degrees from
Beijing Jiaotong University (BJTU), Beijing, China,
in 2004 and 2010, respectively. From April 2010 to
February 2013, he was a Postdoctoral Research Fellow
with the Department of Electrical Engineering,
Tsinghua University, Beijing. Since March 2013, he
has been a Lecturer of BJTU. He is currently an Associate
Professor with the School of Computer and
Information Technology, BJTU. He has published
more than 60 academic papers in referred journals
and conferences. Dr. Xiong serves as a reviewer for the IEEE Transactions
on Wireless Communications, IEEE Transactions
on Communications, IEEE Transactions On Vehicular
Technology, IEEE Wireless Communication Letters
, IEEE ICC,
and IEEE WCNC. He also served as a Session Chair for IEEE GLOBECOM
2012, IET ICWMMN 2013, IEEE ICC 2013, ACM MOMM 2014 and the Publicity and Publication Chair for IEEE
HMWC 2014. His current research interests include
wireless cooperative networks, high-speed railway communications, wireless network coding, and network
information theory. 

\end{biography}

\begin{biography}[]{Pingyi Fan}
received the B.S and M.S. degrees from the Department of Mathematics
of Hebei University in 1985 and Nankai University in 1990,
respectively, received his Ph.D degree from the Department of
Electronic Engineering, Tsinghua University, Beijing, China in 1994.
He is a professor of department of EE of Tsinghua University
currently. From Aug. 1997 to March. 1998, he visited Hong Kong
University of Science and Technology as Research Associate. From
May. 1998 to Oct. 1999, he visited University of Delaware, USA, as
research fellow. In March. 2005, he visited NICT of Japan as
visiting Professor. From June. 2005 to May 2014, he visited Hong
Kong University of Science and Technology for many times and From
July 2011 to Sept. 2011, he is a visiting professor of Institute of
Network Coding, Chinese University of Hong Kong.

Dr. Fan is a senior member of IEEE and an oversea member of IEICE.
He has attended to organize many international conferences including
as General co-Chair of IEEE VTS HMWC2014, TPC co-Chair of IEEE International Conference on Wireless
Communications, Networking and Information Security (WCNIS 2010) and
TPC member of IEEE ICC, Globecom, WCNC, VTC, Inforcom etc. He has served as
an editor of IEEE Transactions on Wireless Communications,
Inderscience International Journal of Ad Hoc and Ubiquitous
Computing and Wiley Journal of Wireless Communication and Mobile
Computing. He is also a reviewer of more than 26 international
Journals including 16 IEEE Journals and 8 EURASIP Journals. He has
received some academic awards, including the IEEE Globecom 2014 Best Paper Award, IEEE WCNC'08 Best Paper
Award, ACM IWCMC'10 Best Paper Award and IEEE ComSoc Excellent
Editor Award for IEEE Transactions on Wireless Communications in
2009. His main research interests include B4G technology in wireless
communications such as MIMO, OFDMA, etc., Network Coding, Network
Information Theory and Cross Layer Design etc.
\end{biography}

\begin{biography}[]{Xianwei Zhou}
received his B.S.
degree in Department of Mathematics
from Southwest Normal
University in 1986 and his M.S.
degree from Zhengzhou University
in 1992, and in 1999, he
obtained Ph.D. degree in Department
of Transportation Engineering
from Southwest Jiaotong
University, China. He was
engaged in postdoctor study
at Beijing Jiaotong University,
China, from 1999 to 2000. Now, he is a full professor in Department
of Communication Engineering,
School of Computer and Communication Engineering, University of
Science and Technology Beijing. He has published more than 100 academic papers in referred journals
and conferences. He won the Baosteel Education Fund outstanding teacher Award and the 2nd Prize of the Nonferrous metallurgy Ministerial level Sci-Tech Advance Award. He was the founding Chair of International Conference on Cyberspace Technology.
His research interests include the security
of communication networks, next generation networks, scheduling
and game theory.
\end{biography}
\end{document}